# Polarization doping – *ab initio* verification of the concept: charge conservation and nonlocality

Ashfaq Ahmad[1], Pawel Strak[1], Pawel Kempisty[1,2], Konrad Sakowski[1,3], Jacek Piechota[1], Yoshihiro Kangawa[2], Izabella Grzegory[1], Michal Leszczynski[1], Zbigniew R. Zytkiewicz,[4] Grzegorz Muziol[1], Eva Monroy[5], Agata Kaminska[1,4,6] and Stanislaw Krukowski[1]*

[1]*Institute of High Pressure Physics, Polish Academy of Sciences, Sokolowska 29/37, 01-142 Warsaw, Poland*

[2]*Research Institute for Applied Mechanics, Kyushu University, Fukuoka 816-8580, Japan*

[3]*Institute of Applied Mathematics and Mechanics, University of Warsaw, 02-097 Warsaw, Poland*

[4]*Institute of Physics, Polish Academy of Sciences, Aleja Lotnikow 32/46, PL-02668 Warsaw, Poland*

[5]*Univ. Grenoble-Alpes, CEA, Grenoble INP, IRIG, PHELIQS , 17 av. des Martyrs, 38000 Grenoble, France*

[6]*Cardinal Stefan Wyszynski University, Faculty of Mathematics and Natural Sciences. School of Exact Sciences, Dewajtis 5, 01-815 Warsaw, Poland*

Abstract

In this work, we study the emergence of polarization doping in $Al_xGa_{1-x}N$ layers with graded composition from a theoretical viewpoint. It is shown that bulk electric charge density emerges in the graded concentration region. The magnitude of the effect, i.e. the relation between the polarization bulk charge density and the concentration gradient is obtained. The appearance of mobile charge was investigated using the combination of *ab initio* and drift-diffusion models. It was shown that the *ab intio* results can be recovered precisely by proper parameterization of drift-diffusion representation of the complex nitride system. It was shown that the mobile charge appears due to the increase of the distance between opposite polarization-induced charges. It was demonstrated that for sufficiently large space distance between polarization charges the opposite mobile charges are induced. We demonstrate that the charge conservation law applies for fixed and mobile charge separately, leading to nonlocal compensation phenomena involving (i) the bulk fixed and polarization sheet charge at the heterointerfaces and (ii) the mobile band and the defect charge. Therefore two charge conservation laws are obeyed that induces nonlocality in the system. The magnitude of the effect allows obtaining technically viable mobile charge density for optoelectronic devices without impurity doping (donors or acceptors). Therefore, it provides an additional tool for the device designer, with the potential to attain high conductivities: high carrier concentrations can be obtained even in materials with high dopant ionization energies, and the mobility is not limited by scattering at ionized impurities.

*Corresponding author, email: stach@unipress.waw.pl

1. Introduction.

There are a number of materials whose crystalline symmetry allows the occurrence of spontaneous polarization, e.g. wurtzite III-nitride semiconductors [1]. The spontaneous polarization is proportional to the shift of the center of mass of electron charge with respect to the atomic core, which leads to the emergence of electric dipoles [1]. In III-nitrides, the magnitude of such dipoles is different for GaN, AlN and InN. Additionally, external strain may change the lattice symmetry, which may, in turn, lead to the appearance of polarization or a change in the polarization [2,3]. Polarization induces a variety of physical effects. In heterostructures, polarization leads to the emergence of electric fields. Generally, for thick layers, the macroscopic electric fields are negligible due to charge screening [3]. Depending on the electric properties the screening may be of different nature, described by Debye-Hückel or Thomas-Fermi approximations [4,5]. Polarization-induced electric fields are observed at nanometer-distance of heterostructures only. An example of a device that can exploit the polarization-induced electric field is the GaN-based field-effect transistor [6]. In other devices, such as laser diodes (LDs) and light-emitting diodes (LEDs) based on III-nitride multi-quantum-well structures, electric fields can play a detrimental role by reducing the radiative recombination probability, which is known as the quantum-confined Stark effect [7-9]. Efforts to remove these fields by fabricating devices on nonpolar crystallographic planes brought only partial success due to the material anisotropy, larger lattice mismatch, and the difficulty to synthesize nonpolar substrates [10].

In devices containing heterostructures, the difference in polarization results in a polarization charge sheet [11,12] and a surface dipole layer [12] at the heterointerface. The former leads to a difference in the electric field at both sides of the heterointerface. The nature of this phenomenon is well understood although the magnitude of the effect in wurtzite nitrides is still under debate [13-15]. The latter leads to an electric potential jump at the interface. The surface dipole effect was identified several decades ago, although its existence, as well as its magnitude, is still disputed [16].

In addition to these well-known effects at heterointerfaces, a different polarization-related phenomenon is predicted to occur in materials with chemical gradients [18-23]. A variation of the concentration in ternary alloys leads to a variation of the polarization, which results in an electrostatic effect that is not encountered in chemically uniform materials. In uniform materials, the electron charge shift in an atom is compensated by the next neighboring atom, thus effective volume charge is zero except at the surface (or heterointerface) where the dipoles are not compensated and create sheets of charge. In a graded material, the dipole changes along the growth axis. Therefore, the net bulk charge density arises, and this phenomenon is denoted as *polarization doping* [18-23].

Assuming that the polarization varies linearly with the alloy concentration, linearly graded materials will contain a uniform non-compensated volume charge density. For nanometer-scale systems, this charge is not screened. In that case, the electrostatic potential distribution may be used to identify the magnitude of the polarization-induced charge density.

In this work, we will verify the existence of the effect by determination of the charge density induced by polarization doping in a graded concentration region via the Poisson equation. The magnitude of the effect will be calculated and a universal scaling constant for the effect, called polarization-doping charge density constant will be determined. We introduce a scaling parameter that can be used to obtain the polarization-doping charge density for any linear concentration gradient. This parameter is directly related to polarization difference, thus it depends on spontaneous polarization and the piezoelectric effects.

2. Methods

*Ab initio* simulations used the package SIESTA, with a numeric atomic orbitals serving as functional base [23,24]. The following atomic orbitals were used: for Al - 3s: DZ (double zeta), 3p: TZ (triple zeta), 3d: DZ; N - 2s: DZ, 2p: TZ, 3d: SZ (single zeta) and for Ga- 4s: DZ (double zeta), 4p: TZ (triple zeta), 3d: DZ. The effective treatment of large systems is enabled by the application of pseudopotentials for all atoms, generated by the ATOM program using all-electron calculations provided by the authors of the code. SIESTA considers norm-conserving Troullier-Martins pseudopotentials, in the Kleinmann-Bylander factorized form [25,26]. Within the Generalized Gradient Approximation (GGA) the PBEsol modification of Perdew, Burke and Ernzerhof exchange-correlation functional was used [27,28].

The lattice constants of bulk AlN, obtained in calculations for a fully periodic system, were $a_{AlN}^{DFT} = 3.112$ Å and $c_{AlN}^{DFT} = 4.983$ Å. These values are in good agreement with the experimental data for wurtzite bulk AlN obtained from x-ray diffraction measurements: $a_{AlN}^{exp} = 3.111$ Å and $c_{AlN}^{exp} = 4.981$ Å [29]. A similar test for GaN gave $a_{GaN}^{DFT} = 3.194$ Å and $c_{GaN}^{DFT} = 5.186$ Å, in good agreement with the experimental values $a_{GaN}^{exp} = 3.1890$ Å and $c_{GaN}^{exp} = 5.1864$ Å [30]. All presented dispersion relations are plotted as obtained from DFT calculations by using Ferreira et al. correction scheme called GGA-1/2 approximation [31]. It includes a self-energy of excitations in semiconductors, providing bandgap (BG) energies, effective masses, and band structures in good agreement with experiments.[32] Our results were obtained using GGA-1/2 correction scheme to structures in which positions of atoms and a periodic cell were first relaxed to equilibrium state by using GGA-PBEsol approximation. In geometry optimizations, all atoms were relaxed to reduce the forces on single atoms below 0.005 eV/Å. Spin-orbit effects were neglected in present calculations since the high-lying valence and low-lying conduction states lead to only a small amount of splitting (on the order of 10 meV). The bandgap of bulk AlN was then equal: $E_g^{DFT}(AlN) = 6.19\ eV$ in good agreement with the experimental data of Silveira et al. ($E_g^{exp}(AlN) = 6.09\ eV$) [33]. Similarly, values for GaN were $E_g^{DFT}(GaN) = 3.47\ eV$ and $E_g^{exp}(GaN) = 3.47 eV$, respectively [34,35]. As a convergence criterion to terminate a self-consistent field (SCF) loop, the maximum difference between the output and the input of each element of the density matrix had to be equal or smaller than $10^{-4}$.

The device part was simulated within the drift-diffusion model [36,37], using the package developed by the authors [38]. The method solves Van Roosbroeck set of equations (drift-diffusion equations) for the electric potential and the Fermi quasi-levels for electrons and holes. These equations describe the real space dependence of classical fields that are converted into the finite matrix representation of the solution using the discontinuous Galerkin method [39,40]. The set of highly nonlinear coupled matrix equations is solved by a massive self-consistent iteration technique [41]. The solutions provide 2-d and 3-d approximations of all relevant classic fields within devices, such as electric potential, density and drift velocity of the electrons and holes, density of ionized point defects, donors and acceptors. Thus, the micro- and nano-scale picture of the device under equilibrium and non-equilibrium conditions is obtained. The system requires input describing physical properties of all materials, which may be affected by local fields, such as temperature or strains. The latter gives rise to polarization effects, spontaneous and piezo-electric inducing electric fields which are represented by sheet and bulk polarization charge density. These quantities are obtained from the above described *ab initio* calculations.

## 3. Results
### 3.1 *Ab initio* simulations of the graded Al-Ga-N system.

In the *ab initio* calculations, we use graded-$Al_xGa_{1-x}N$/AlN/GaN and graded-$Al_xGa_{1-x}N$/GaN/AlN superlattices. The schematic view of a single period of the two structures is presented in Fig. 1. In the diagrams, the [000-1] direction points towards the top. From bottom to top, the periods consist of:

(a) 15 atomic layers of a linearly graded ternary alloy with Al content increasing along [000-1] direction, 5 atomic layers of pure AlN and 4 atomic layers of pure GaN,

(b) 15 atomic layers of a linearly graded ternary alloy with Al content decreasing along [000-1] direction, 4 atomic layers of pure GaN and 5 atomic layers of pure AlN.

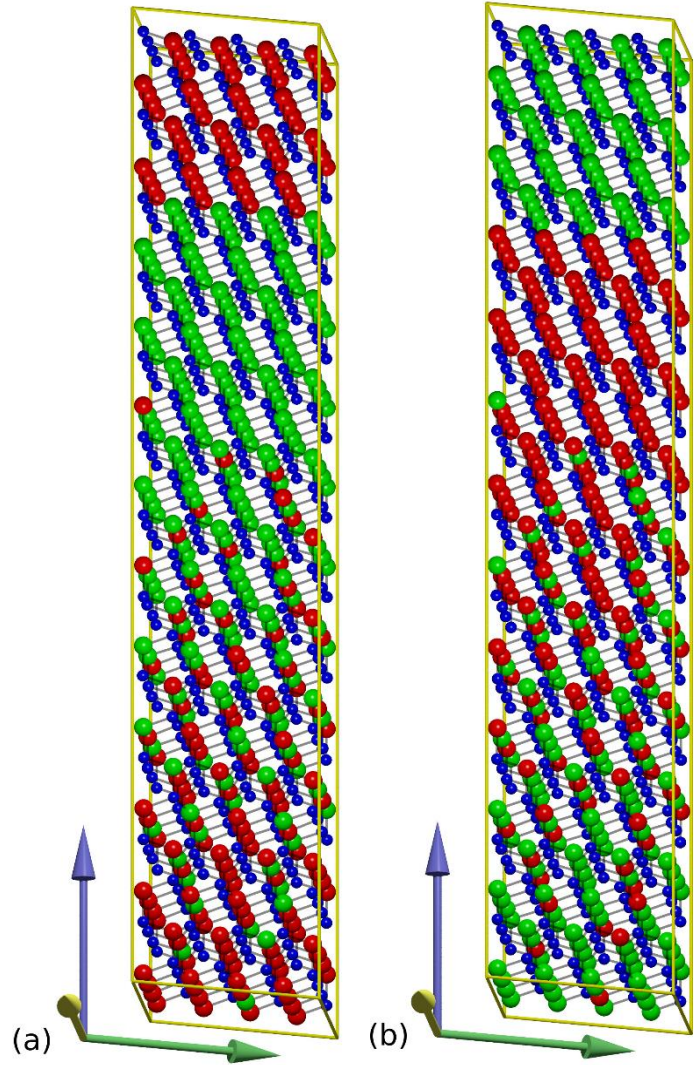

**Fig. 1.** (a) $Al_xGa_{1-x}N/AlN/GaN$ and (b) $Al_xGa_{1-x}N/GaN/AlN$ supercells used in *ab initio* calculations. The $Al_xGa_{1-x}N$ layer is linearly graded, with (a) increasing and (b) decreasing Al content towards the top of the cell ([000-1] direction). Al, Ga and N atoms are represented by green, red and blue balls, respectively.

In the ternary alloy section, the metal atoms in each layer are distributed randomly, and the number of Al atoms in subsequent layers is changed consecutively by one. Thus the number of the Al and Ga atoms is strictly determined within the layer and changes linearly along the z-axis. The lattice positions of Al and Ga atoms within the layer are selected using a random number generator typical for Monte Carlo simulations. Thus the procedure uses uniform sampling in the occupation space, immune to systematic errors but prone to high noise. As shown by potential profiles presented in Fig. 3, the noise is negligible due to long range of Coulomb interactions which averages out the potential profile which is insensitive to the local variation in AlGaN composition. Also, the positions in the neighboring layers are not correlated. Several different configurations were used with no discernible change of the averaged quantities.

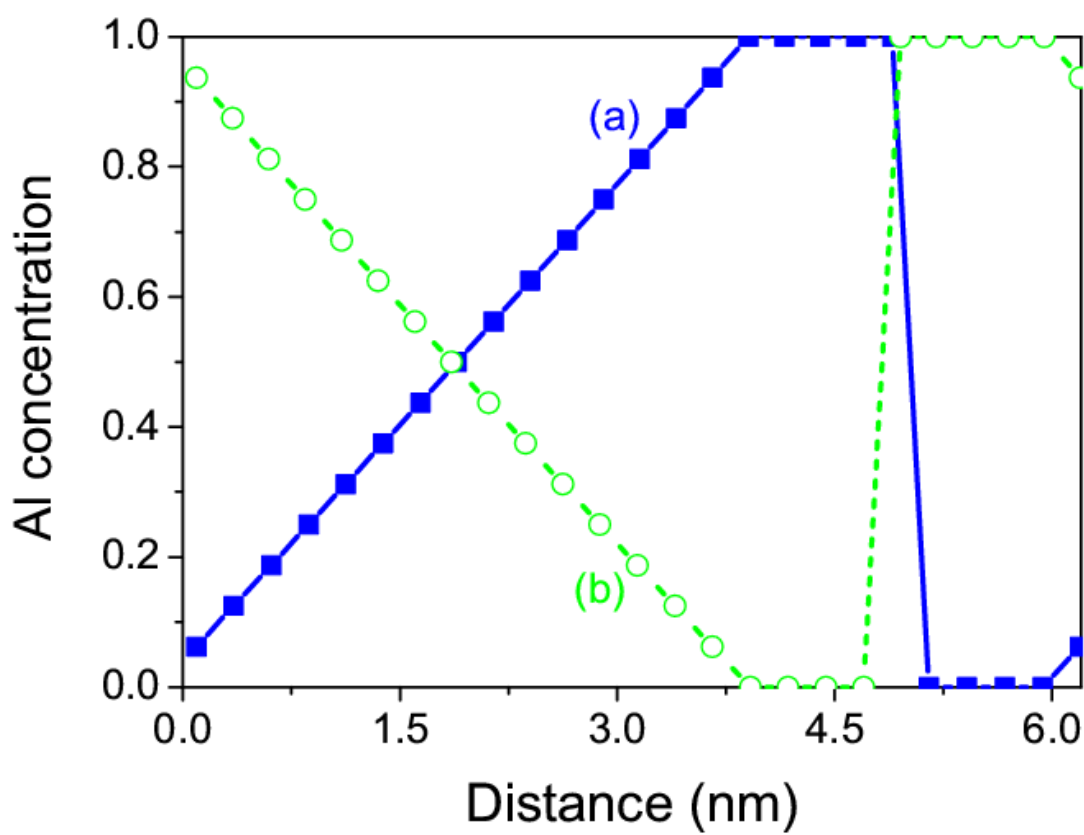

**Fig 2**. Variation of the aluminum concentration (x) along the [000-1] axis in the structures described in Figure 1. The blue solid line, with full squares and the green solid line with empty circles, represent the Al concentration in the lattice shown in Figs 1(a) and 1(b), respectively.

The change in the average composition of the atomic layers is presented in Fig. 2. A 4 × 4 supercell was used. Therefore 16 mixed layers were used and the concentration change at a single layer was $\Delta x = \frac{1}{16} = 0.0625$. This change occurred at the distance of $c/2 = 0.2593 nm$. Therefore the concentration slope is $sc = \frac{dx}{dz} = 0.241 nm^{-1}$. Accordingly, the average local polarization is expected to change linearly with the metal concentration. The linear polarization change should give a volume charge density $\rho(\vec{r})$, which is related to the curvature of electrical potential $\varphi(\vec{r})$ via Poisson equation:

$$\nabla^2 \varphi(\vec{r}) = -\rho(\vec{r})/\varepsilon_b \varepsilon_o \qquad , \qquad (1)$$

where $\varepsilon_b$ is the static dielectric constant of the semiconductor and $\varepsilon_o$ dielectric permittivity. As the dielectric constants of AlN and GaN are approximately equal, the average constant value corresponding to $x = 0.5$ is used ($\varepsilon_b = 10.295$). The electric potential profile obtained by double average procedure from the potential obtained from *ab initio* calculations as described in Ref 16. The potential function, resulting from the solution of Poisson equation in *ab initio* procedure is represented by a finite set of values on the grid in real space. First, the potential is averaged in a plane, to obtain a one-dimensional grid approximation potential function. This potential is still wildly oscillating due to atomic cores contribution. In the second stage, the potential is smoothed by adjacent averaging over the c lattice parameter period. The value is associated with the middle point, and the period is shifted by a single point and the procedure is repeated. The potential is still slightly oscillating therefore adjacent averaging is repeated twice using GaN and AlN c lattice parameters. The results are the potential profiles shown in Fig. 3. The segments corresponding to pure AlN, pure GaN and linearly graded Al$_x$Ga$_{1-x}$N have different functional profiles. The sections with uniform concentration present a linear variation of the potential, whereas the potential profile within the linearly graded alloy can be approximated by a parabola, which is characteristic of systems with uniform bulk charge density. In addition, at the AlN-GaN interface, a potential jump is observed, due to the existence of a dipole layer.

The parabolic potential profile in the graded alloys of (a) and (b) cases can be described by parabolic functions $V_a(z) = a_0 + a_1 z + a_2 z^2$ and $V_b(z) = b_0 + b_1 z + b_2 z^2$, respectively, with $a_0 = 13.627 \pm 0.003 V$, $a_1 = -0.067 \pm 0.003 V/nm$ and $a_2 = 0.14 \pm 0.01 V/nm^2$ for case (a), and $b_0 = 11.971 \pm 0.002 V$, $b_1 = 0.126 \pm 0.002 V/nm$ and $b_2 = -0.16 \pm 0.01 V/nm^2$ for case (b). From these data, the bulk charge density derived via Eq. 1 is: (a) $\rho_{pd} = -2.54 \pm 0.18 \times 10^7 C/m^3$ and (b) $\rho_{pd} = 2.91 \pm 0.18 \times 10^7 C/m^3$.

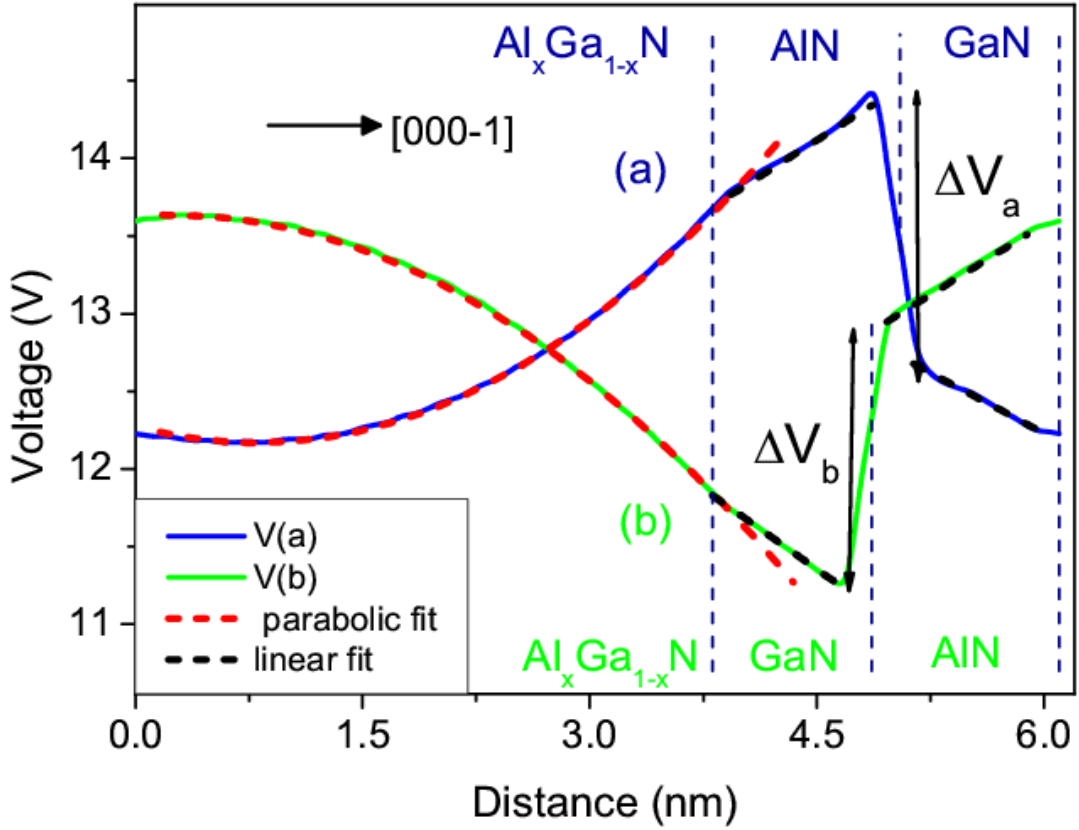

**Fig. 3.** The electric potential profile obtained from *ab initio* calculations. Solid green and blue lines: averaged potential profiles in the structures in Fig.1 (a) and (b), respectively; red dashed lines: parabolic approximation in linear regions; black dashed lines: linear approximation in uniform regions; $\Delta V$: potential jumps at sharp heterointerfaces due to dipole layers.

The above-listed charge densities refer to fixed charges. The same total amount of fixed polarization-induced surface sheet charge at AlN/GaN interface may be estimated invoking the charge conservation law and using the thickness of the graded layer: (a) $d_{grad} \cong 35.6$ Å and (b) $d_{grad} \cong 35.1$ Å. Thus, the obtained surface sheet charge densities are (a) $\rho_{surf} = 9.07 \times 10^{-2} \, C/m^2$ and (b) $\rho_s = -10.21 \times 10^{-2} \, C/m^2$.

Note that the simulated systems are heavily strained, which drastically affects polarization charges via piezoelectric effects [2,13,14,42-45]. As it was demonstrated recently, the piezoelectric contribution may amount to 50% of the polarization induced fields in nitride mulita-quantum well (MQW) [45]. Thus piezoelectric effect plays important role and should be taken into account in the AlN-GaN graded structures. The AlN-GaN polarization difference ($\Delta P_{AlN-GaN} \equiv P_{AlN} - P_{GaN}$) under different strain states is listed in Table 1. The strain was calculated using standard formula $\varepsilon_{11}(GaN) = \frac{a-a_o(GaN)}{a_o(GaN)}$ and $\varepsilon_{11}(AlN) = \frac{a-a_o(AlN)}{a_o(AlN)}$ where as reference values the ab initio lattice parameters were used: $a_o(GaN) = a_{GaN}^{DFT} = 3.194$ Å and $a_o(AlN) = a_{AlN}^{DFT} = 3.112$ Å. In the table, the "Relaxed" case is obtained as the spontaneous polarization difference between bulk AlN and GaN. The lattice constant in the system referred to as "Model strained" was obtained by elastic relaxation of the lattice in the *ab initio* procedure. In this case, the in-plane lattice constant evolved to $a = 3.13434$ Å and the strain along the c axis was $\varepsilon_{33} = -2 C_{13}\varepsilon_{11}/C_{33}$ , where $\varepsilon_{11}$ is the in-plane strain and $C_{ij}$ are elastic constants, obtained from Mahata et al. [46]. These results show that the strain drastically affects the polarization and consequently sheet charge density at the AlN-GaN interface.

Table 1. AlN-GaN polarization difference $\Delta P_{AlN-GaN}$ (considering spontaneous and piezoelectric polarization) and polarization charge density parameter $Q_{pol} = \Delta P_{AlN-GaN}/e$ under various strain states (data from Bernardini, Fiorentini and Vanderbilt [13])

| Strain state \property | $\varepsilon_{11}(GaN)$ | $\varepsilon_{11}(AlN)$ | $\Delta P_{AlN-GaN}$ $(C/m^2)$ | $Q_{pol}$ $(nm^{-2})$ |
|---|---|---|---|---|
| Relaxed | 0 | 0 | -0.052 | -0.324 |
| Strained to GaN | 0 | 0.0264 | -0.106 | -0.661 |
| Strained to AlN | -0.0257 | 0 | -0.087 | -0.544 |
| Model strained | -0.0187 | 0.0072 | -0.092 | -0.575 |
| Directly from model (a) | -0.0187 | 0.0072 | -0.091 | -0.577 |
| Directly from model (b) | -0.0187 | 0.0072 | -0.102 | -0.637 |

The polarization doping charge is compensated by the sheet charge density at the AlN-GaN interface arising from the polarization density difference. The electric fields in the chemically uniform regions of the lattice (a) are: $\vec{E}_{AlN} = -\nabla V = -0.072 \pm 0.002 \frac{V}{Å}$ and $\vec{E}_{GaN} = -\nabla V = 0.062 \pm 0.001 \frac{V}{Å}$, whereas in the lattice (b) the electric fields are: $\vec{E}_{GaN} = -\nabla V = -0.069 \pm 0.001 \frac{V}{Å}$ and $\vec{E}_{AlN} = -\nabla V = 0.061 \pm 0.001 \frac{V}{Å}$. That defines the sheet charge density at the interfaces via Gauss law, in accordance with the equation

$$\rho_s = \varepsilon_o(\varepsilon_1 \vec{E_1} - \varepsilon_2 \vec{E_2}) = \vec{P_2} - \vec{P_1} \qquad (2)$$

that gives (a) $\rho_s = 1.13 \pm 0.02 \times 10^{-1} C/m^2$ and (b) $\rho_s = -1.18 \pm 0.02 \times 10^{-1} C/m^2$. As indicated in Eq. 2, these charges arise from the polarization discontinuity at the AlN-GaN interface. The obtained values are in reasonable agreement (but with opposite sign) with those obtained by integrating the charge density in the graded layers. Therefore, the polarization doping charge density can be related to the bulk polarization difference using the following relation:

$$\rho_{pd} = \Delta P \left(\frac{dx}{dz}\right) \tag{3}$$

where the concentration x is expressed in atomic fractions. In the drift-diffusion device modeling different units are used, employing elementary charge based unit of charge and nm as unit length. The fixed charge may be therefore expressed in terms of the density of carriers (n - electrons or p - holes):

$$p, n = \frac{\rho_{pd}}{e} = \left(\frac{\Delta P}{e}\right)\left(\frac{dx}{dz}\right) = Q_{pol}\left(\frac{dx}{dz}\right) \tag{4}$$

where polarization charge density $\rho_{pd}$ is determined via parameter $Q_{pol}$ that is expressed in $nm^{-2}$. Thus, the polarization doping charge may be characterized by polarization charge density constant

$$Q_{pol} = \frac{\Delta P}{e} = \frac{\rho_{pd}}{e\left(\frac{dx}{dz}\right)} \tag{5}$$

As compensation by defects and screening are not included in these equations, the above values set an upper limit for polarization-induced mobile charge density as it will be discussed below.

Further verification of the model was obtained by calculation of a system with reduced concentration slope by a factor of 2 i.e. by the replacement of a single Al atom by Ga atom in every second layer. For the same graded layer thickness, the concentration slope was reduced to $\frac{dx}{dz} = 0.120 nm^{-1}$. The thickness of the uniform region was also the same as that used previously. This opens the possibility of calculations of the system with a reduced Al concentration change. Thus we consider (i) a graded-$Al_xGa_{1-x}N/Al_{0.5}Ga_{0.5}N/GaN/$ (15/5/4 atomic layers) periodic heterostructure with x varying from 0 to 0.5 along the [000-1] axis, and (ii) a graded-$Al_xGa_{1-x}N/AlN/Al_{0.5}Ga_{0.5}N$ (15/5/4 atomic layers) periodic heterostructure with x varying from 0.5 to 1 along the [000-1] axis. The obtained parabolic coefficients were $a_2(2) = 0.073 \pm 0.01 V/nm^2$ and $a_2(2') = 0.065 \pm 0.01 V/nm^2$ for structures (i) and (ii), respectively. The polarization charge density parameters are $Q_{pol}(2) = -0.688 \pm 0.094 nm^{-2}$ and $Q_{pol}(2') = -0.613 \pm 0.094 nm^{-2}$, reasonably close to previously obtained values $Q_{pol}(a) = -0.577 nm^{-2}$ and $Q_{pol}(b) = -0.637 nm^{-2}$. For the inverted concentration Al gradients, (i) extending from x = 1.0 to 0.5 and (ii) from x = 0.5 to 0.0, the parabolic coefficients were $b_2(2) = 0.064 \pm$ $0.01\, V/nm^2$ and $b_2(2') = 0.060 \pm 0.01\, V/nm^2$. The derived bulk charge density are $\rho(2) = 1.16 \pm 0.18 \times 10^7\, \frac{C}{m^3}$ and $\rho(2') = 1.09 \pm 0.18 \times 10^7\, \frac{C}{m^3}$. The polarization charge density parameters are: $Q_{pol}(2) = 0.603 \pm 0.094\, nm^{-2}$ and $Q_{pol}(2') = 0.565 \pm 0.094\, nm^{-2}$.

The obtained data may be compared with experimental values. For instance, Zhu et al obtained $Q_{pol}^{exp} = 0.3\, nm^{-2}$ [47], Yan et al reported $Q_{pol}^{exp} = 0.2\, nm^{-2}$ [48], Amstrong et al measured $Q_{pol}^{exp} = 0.48\, nm^{-2}$ [49], and Li et al. obtained $Q_{pol}^{exp} = 0.33\, nm^{-2}$ [50]. These experimental values are lower than those theoretically derived. Experimentally, the polarization-induced fixed charge is not measured directly but estimated from the resulting carrier concentration, since only mobile charge contributes to electric conductivity. As will be shown below, the polarization charge density values obtained from DFT measurements are always higher than the density of mobile charge. The agreement between the data from the model and the direct experimental data may be assessed as reasonably good.

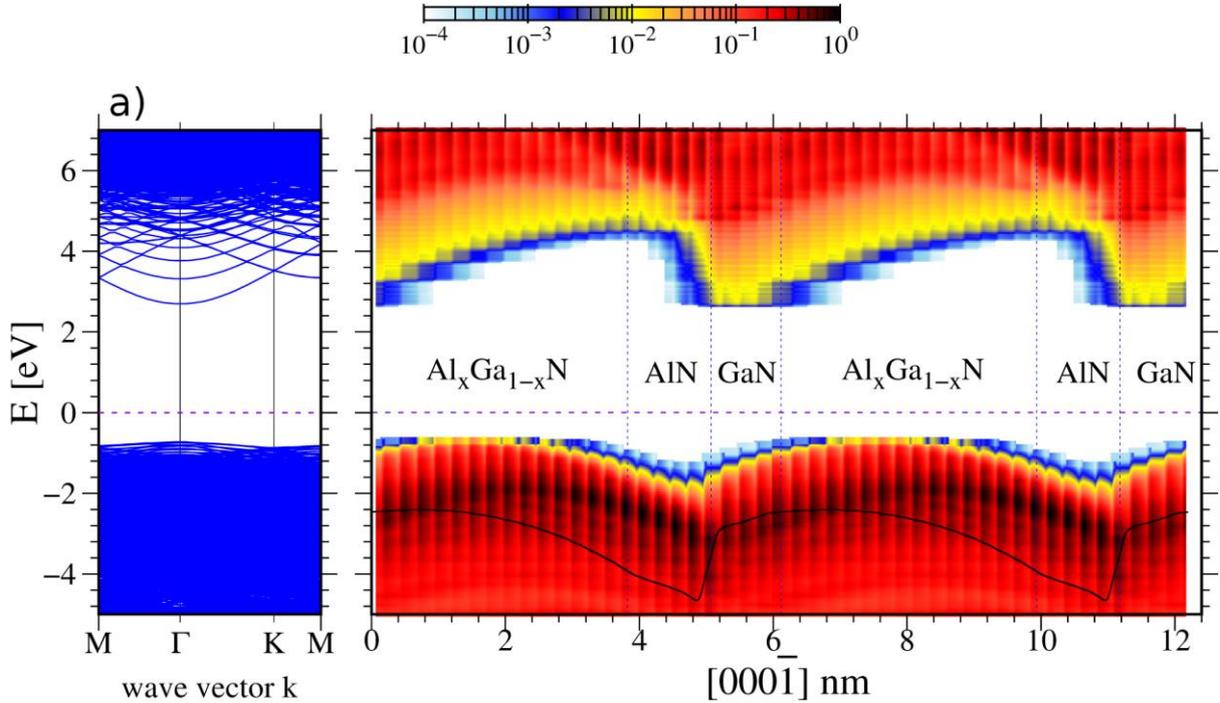

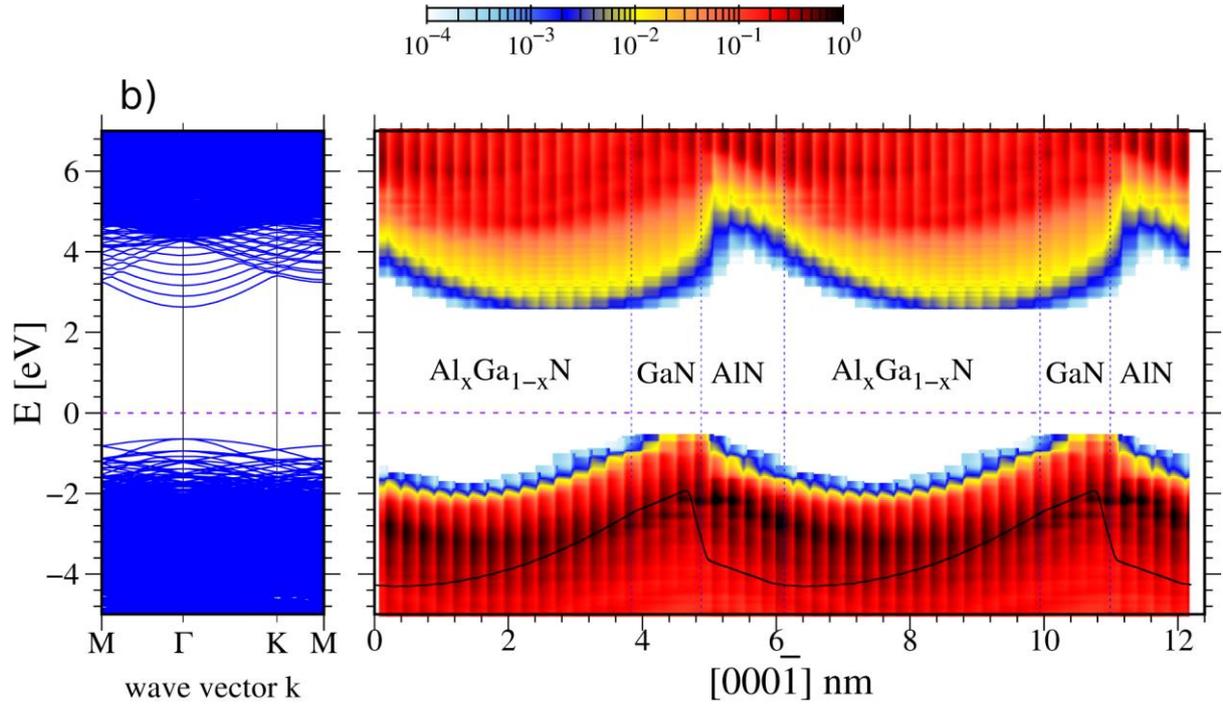

**Fig. 4.** Band profiles where (a) and (b) denote lattices presented in Fig. 1 (a) and (b) respectively. The diagram presents: left panes – bands in momentum space, right panel – bands in coordination space. The color scale at the top denotes the density of states. The black line superimposed on the valence band represents the electric potential multiplied by the electron charge.

Finally, the graded-$Al_xGa_{1-x}N$/AlN/GaN and graded-$Al_xGa_{1-x}N$/GaN/AlN periodic structures described in Figs 1-3, have band profiles calculated from ab initio data by a projection of band states on atom rows. The procedure was used previously both for AlN/GaN MQWs [16] and surface slab [51] simulations. The results, presented in Fig. 4, prove that the Fermi level is located deep in the bandgap so that the contribution of free carriers can be neglected. Therefore determination of the polarization charge from the electric potential profile is positively verified. An important subject of the emergence of the mobile charge in polarization-doping system is presented in the following Section. The effect has large technological potential as scattering by ionized defects is not present [49,52]. Thus the alloy scattering is the dominant mechanism limiting carrier mobility [19,51,53].

**3.2 Drift-diffusion simulation of the emergence of the mobile charge.**

The emergence of the mobile charge requires much larger system sizes than those previously described. Therefore drift-diffusion simulations were used [36-41]. Initially, the system employed DFT simulations presented in Fig. 1 was calculated. The results are presented in Fig. 5.

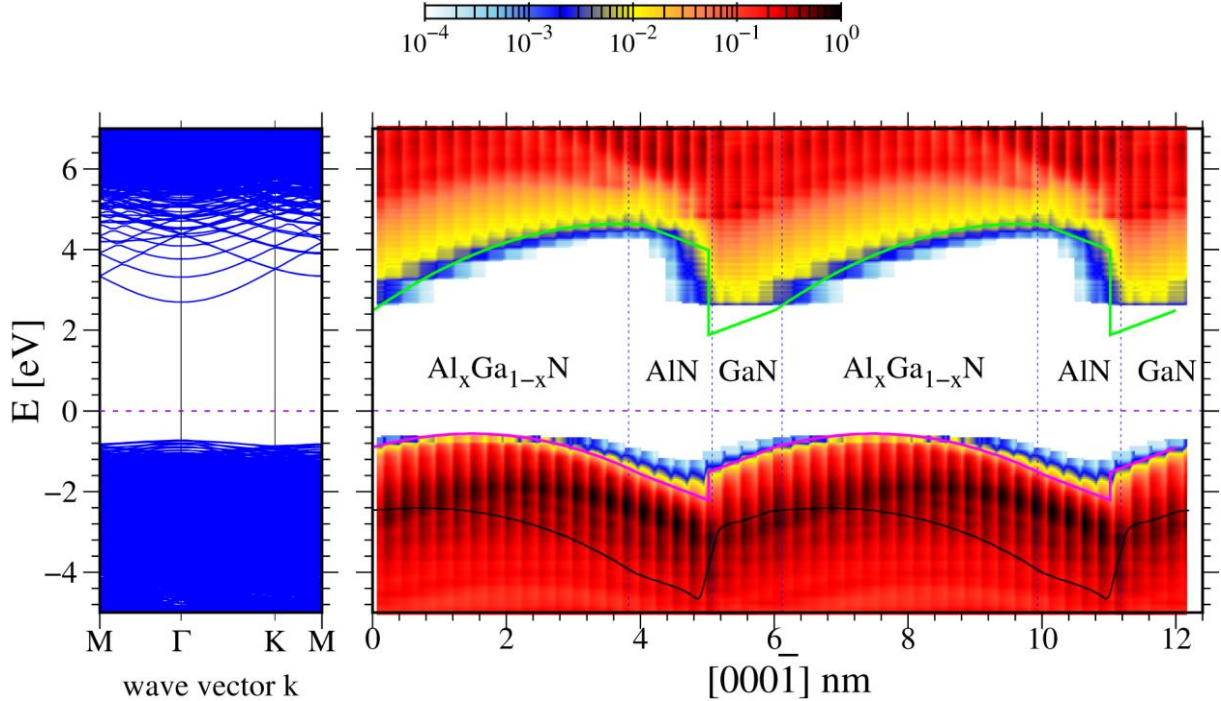

**Fig. 5.** Comparison of the *ab initio* and drift-diffusion simulations of the system presented in Fig. 1 case (a). The color map presents the top denotes the density of states obtained from *ab initio* simulations. The black line superimposed on the valence band represents the electric potential multiplied by the electron charge. The red line represent the edges of the valence and conduction band obtained from drift-diffusion calculations.

The drift-diffusion band profiles are essentially identical to those obtained by *ab initio* calculations. Thus the essential issue of compatibility of the ab initio and drift-diffusion models is solved.

The next step is to increase the distance between the graded AlGaN region and the GaN/AlN interface, i.e. to extend GaN regions, as shown in Fig. 6. As expected the potential slope of the GaN regions is preserved, so that the Fermi level is adjacent to both bands. As is expected the Fermi level entry into the valence band (VB) and conduction band (CB) occurs simultaneously, because the polarization and mobile charge have to be conserved separately. The diagram illustrates the fundamental principle of the polarization-doping region that the mobile charge emerges due to increasing spatial separation of the positive and negative polarization charge.

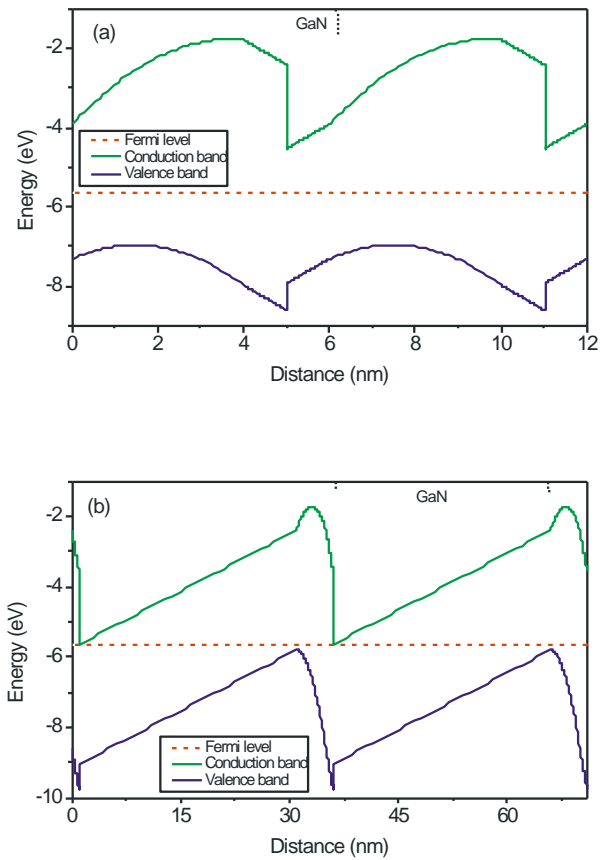

**Fig. 6.** Influence of polarization-doping charge spatial separation elucidated by drift-diffusion simulations: (a) system presented in Fig 5, (b) system with the thickness of GaN layer increased.

As it is shown in Fig. 6, the increase of the thickness of GaN layer, equivalent to the increase of the separation of polarization-doping positive and negative charges preserves the electric field in the GaN layer. The effect is analogous to capacitor, where two parallel infinite charges induce the uniform electric field. As the Fermi level is deep in the bandgap, the mobile charges are absent. Finally, the Fermi level touches and penetrates the conduction and valence band which leads to emergence of mobile charge. Therefore the route to the creation of mobile charge leads by the separation of the polarization charges. This route will be used in the design of the UV device below.

Finally, the entirely new design of the doping-free light emitting diode is presented. The design may be particularly useful for deep UV device as there is no shallow impurities for n-and p-type in Al rich nitrides. The results presented in Fig. 7 confirm those in Fig. 6. The presence of p-n junction and quantum wells does not affect the principles formulated above: (i) the total polarization charge is compensated to zero in the entire system (ii) the p-n junction does not change the general trend according to which the mobile charge emerges due to the increase of the distance between positive and negative polarization charge.

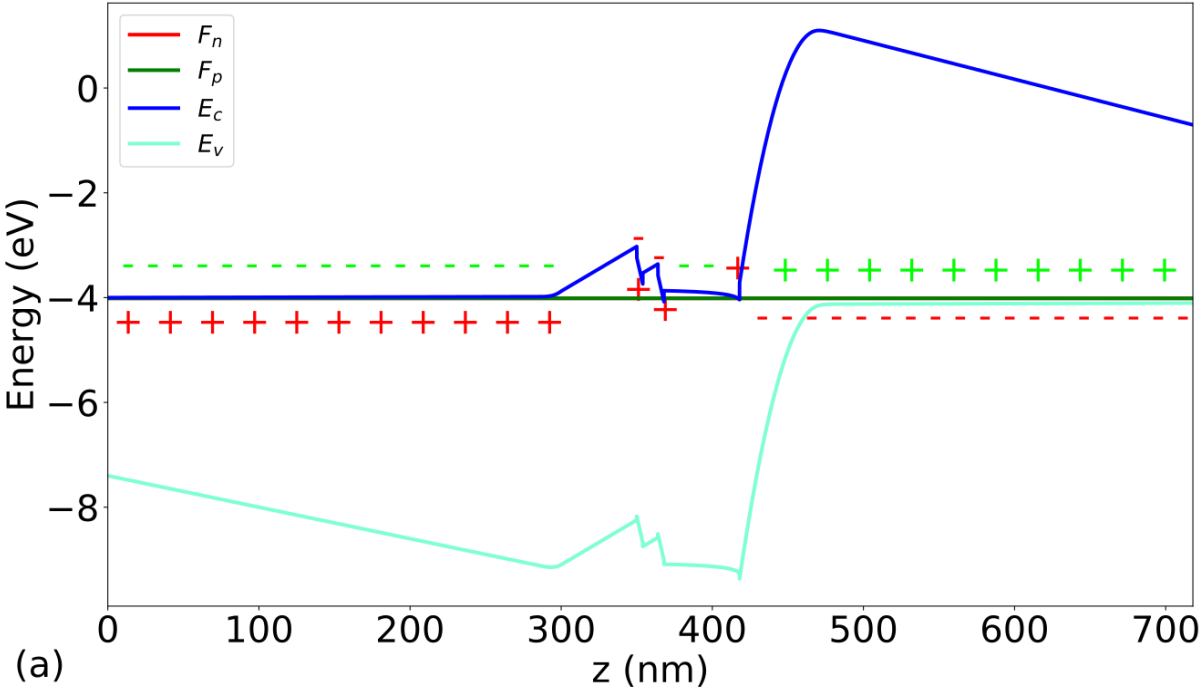

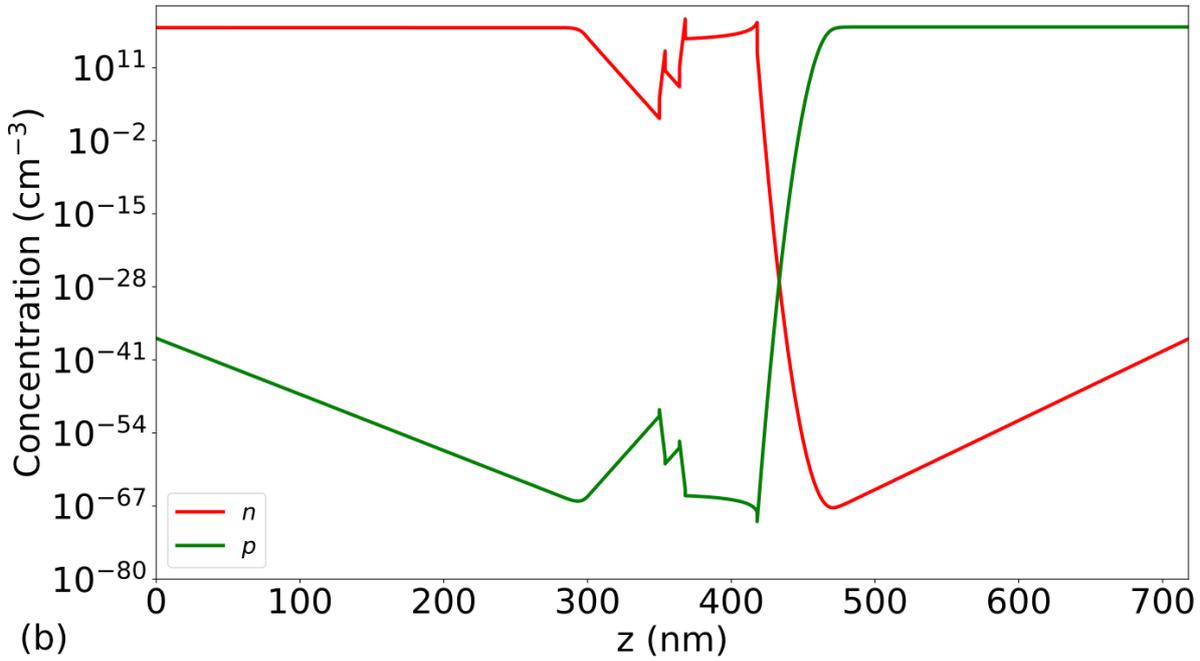

Fig. 7. Drift-diffusion simulations of doping-free LED with the graded zone in the n- and p-type regions : (a) band profile, (b) mobile charge density. Red, green and blue symbols mark the polarization, mobile and defect charge, respectively. Notation: n, p – electron and hole concentration; $F_n$, $F_p$ – electron and hole quasi-Fermi levels (equal); $E_c$, $E_v$ – conduction (valence) band minimum (maximum).

The opposite charge is not larger than the polarization charge. In general, for larger distances, the charge is larger. This observation is important for the comparison of the previously derived values $Q_{pol}(a) = -0.577 nm^{-2}$ and $Q_{pol}(b) = -0.637 nm^{-2}$. For the reduced gradients these values were: $Q_{pol}(2) = -0.688 \pm 0.094 nm^{-2}$ a $Q_{pol}(2') = -0.613 \pm 0.094 nm^{-2}$ and $Q_{pol}(2) = 0.603 \pm 0.094 nm^{-2}$ and $Q_{pol}(2') = 0.565 \pm 0.094 nm^{-2}$. The obtained data may be compared with experimental values. For instance, Zhu et al obtained $Q_{pol}^{exp} = 0.3 nm^{-2}$ [47], Yan et al reported $Q_{pol}^{exp} = 0.2 nm^{-2}$ [48], Amstrong et al measured $Q_{pol}^{exp} = 0.48 nm^{-2}$ [49], and Li et al. obtained $Q_{pol}^{exp} = 0.33 nm^{-2}$ [50]. These experimental values are significantly lower than those theoretically derived. From the above analysis, it follows that the theoretically achieved values may be attained only for the infinite distance between polarization charges. Taking into account this argument it is appropriate to conclude that these experimental data confirm theoretically derived values.

4. **Summary**

This work provides atomic-scale model and *ab initio* proof of the existence of polarization doping in $Al_xGa_{1-x}N$ ternary alloys with graded concentration. It has been shown that in linearly graded regions, a constant bulk volume charge arise. The obtained charge density, at extremely high concentration gradients of $0.241 nm^{-1}$, is $n, p > 10^{20} cm^{-3}$, which corresponds to doping levels far above the Mott transition in III-nitride materials. Attaining this doping level with doping impurities implies a reduction of the mobility due to scattering by ionized impurities, and most often a major degradation of the structural quality of the material.

It is also shown that charge conservation and charge neutrality leads to highly nonlocal phenomena, which could be denoted as device-size compensation. The fixed charge neutrality links the polarization doping bulk charge in $Al_xGa_{1-x}N$ graded layer with the polarization charge at the AlN/GaN interface. Thus it is subject to strong piezoelectric effects. For an AlN/GaN strained device, the charge related to the difference in polarization is $\Delta P_{AlN-GaN} = 0.92 (C/m^2)$, and our calculations predict that the charge derived from polarization doping is $\Delta P_{AlN-GaN} = 0.91 (C/m^2)$, which confirms this identification.

The appearance of mobile charge was investigated using the combination of *ab initio* and drift-diffusion models. It was shown that the *ab intio* results can be recovered precisely by proper parameterization of drift-diffusion representation of the complex nitride system. It was shown that for sufficiently small system size mobile charge is not present. The increase of the distance between positive and negative polarization charges leads to the Fermi level entry into the conduction and valence bands simultaneously. Thus the polarization and mobile charge is conserved separately. The latter charge may have a contribution from doping, which affects mobile charge in a standard way. Therefore for the existence of the mobile charge, the separation of the positive and negative polarization charge should be sufficient.

**Acknowledgements**

The research was partially supported by Poland National Centre for Research and Development [grant number: TECHMATSTRATEG-III/0003/2019-00] and partially by Japan JST CREST [grant number JPMJCR16N2] and by JSPS KAKENHI [grant number JP16H06418]. This research was carried out with the support of the Interdisciplinary Centre for Mathematical and Computational Modelling at the University of Warsaw (ICM UW) under grants no GB77-29, GB84-23 and GB76-25.